\documentclass{WileyChemistry-template}

\title{\raggedright Doping-Induced Enhancement of Hydrogen Evolution at \ce{MoS2} Electrodes}

\author{
\begin{minipage}{\textwidth}	
	Sander \O. Hanslin,\textsuperscript{[a,b]} Hannes J\'onsson,*\textsuperscript{[b]} Jaakko Akola,*\textsuperscript{[a,c]}
\end{minipage}
}

\newcommand{\affiliation}{
\begin{itemize}	    

\item[{[a]}] 
Department of Physics, Norwegian University of Science and Technology, NO-7491 Trondheim, Norway\\

\item[{[b]}] 
Science Institute and Faculty of Physical Sciences, University of Iceland, IS-107 Reykjav{\'\i}k, Iceland\\

\item[{[c]}] 
Computational Physics Laboratory, Tampere University, FI-33101 Tampere, Finland\\

E-mail: hj@hi.is,\ jaakko.akola@ntnu.no\\
\end{itemize}
}

\renewcommand{\abstract}{Rate theory and DFT calculations of hydrogen evolution reaction (HER) on \ce{MoS2}  with \ce{Co}, \ce{Ni} and \ce{Pt} impurities show the significance of dihydrogen (\ce{H2}*) complex where both hydrogen atoms are interacting with the surface. Stabilization of such a complex affects the competing Volmer-Heyrovsky (direct \ce{H2} release) and Volmer-Tafel (\ce{H2}* intermediate) pathways. The resulting evolution proceeds with a very small overpotential for all dopants ($\eta = 0.1$ to $0.2$~V) at $25$\,\% edge substitution, significantly reduced from the already low $\eta = 0.27$~V for the undoped edge. At full edge substitution, \ce{Co}-\ce{MoS2} remains highly active ($\eta = 0.18$~V) while \ce{Ni}- and \ce{Pt}-\ce{MoS2} are deactivated ($\eta = 0.4$ to $0.5$~V) due to unfavorable interaction with \ce{H2}*. Instead of the single \ce{S}-vacancy, the site of intrinsic activity in the basal plane was found to be the undercoordinated central \ce{Mo}-atom in threefold \ce{S}-vacancy configurations, enabling hydrogen evolution with $\eta=0.52$~V via a \ce{H2}* intermediate. The impurity atoms interact favorably with the intrinsic sulfur vacancies on the basal plane, stabilizing but simultaneously deactivating the triple vacancy configuration. The calculated shifts in overpotential are consistent with reported measurements, and the dependence on doping level may explain variations in experimental observations. 
}

\newcommand{\keywords}{
	Computational electrocatalysis \textbullet\
	Grand canonical DFT \textbullet\  
	Implicit solvation \textbullet\ 
    Hydrogen evolution \textbullet\ 
	Molybdenum disulfide
}

\begin{document}
\emergencystretch 1em

\twocolumn[\vspace{-1.5cm}\maketitle\vspace{-1cm}
	\textit{\dedication}\vspace{0.4cm}]
\small{\begin{shaded}
		\noindent\abstract
	\end{shaded}
}

\begin{figure} [!b]
\begin{minipage}[t]{\columnwidth}{\rule{\columnwidth}{1pt}\footnotesize{\textsf{\affiliation}}}\end{minipage}
\end{figure}



\section*{Introduction}
\label{introduction}

\ce{MoS2} has emerged as a promising candidate among earth-abundant compounds to replace the precious metal catalysts traditionally used for the hydrogen evolution reaction (HER) \cite{doi:10.1021/ja0504690,doi:10.1126/science.1141483,B803857K,Kibsgaard2012,doi:10.1021/cs300451q,https://doi.org/10.1002/adma.201302685,https://doi.org/10.1002/chem.201500435}. Its two-dimensional layered nature allows for novel engineering on the nanoscale, for example by maximizing the presence of edge sites, which display higher activity than the basal plane. 

Transition metal doping has been successful in improving the reaction rates \cite{Wang2015,C5EE00751H,doi:10.1021/acsami.5b08420,C6EE01786J,doi:10.1021/acscatal.6b01274,doi:10.1021/jacs.7b08881,C8SC01114A,surfaces2040039,Zheng2020,doi:10.1021/acsaem.1c01721,SUNDARAVENKATESH202237256}, but different and sometimes conflicting results raise interesting questions regarding the underlying activation mechanism. For example, Deng et al. \cite{C5EE00751H} found significant reduction of the HER overpotential with \ce{Pt}-doping of few-layer \ce{MoS2} samples, while \ce{Co} and \ce{Ni} showed signs of weak activation and deactivation, respectively. Lau et al. \cite{C8SC01114A} found that \ce{Co}-doping lead to activation with respect to undoped \ce{MoS2}, while \ce{Fe}, \ce{Ni}, and \ce{Ag} were detrimental for the electrochemical current density. Wang et al. \cite{Wang2015} found reduction in the HER overpotential for \ce{Fe}, \ce{Co}, \ce{Ni}, \ce{Cu} in vertically aligned \ce{MoS2}, while Humphrey et al. \cite{C9NR10702A} conversely found that \ce{Co}-doped \ce{MoS2} displayed a larger overpotential on basal-oriented \ce{MoS2}. 

In light of these rich properties of doped \ce{MoS2}, it is of interest to determine which mechanisms are responsible for the experimentally observed activation or deactivation. The experimental situation is complex, and the doping, especially at high levels, may lead to significant changes in the sample morphology, or other large-scale modifications such as phase transitions, phase separation or formation of impurity particles or clusters on the \ce{MoS2} substrate. Such effects are relevant for catalyst performance, but constitute a regime which is challenging to assess theoretically due to the wide scope. In this work, we therefore limit our focus to the relatively low concentration doping regime, considering single- or few-atom impurities in the 2H phase of \ce{MoS2}. We note that multi-elemental codoping is also a promising approach towards increasing \ce{MoS2} activity \cite{NGUYEN2021105750,https://doi.org/10.1002/adma.202001167,doi:10.1021/acssuschemeng.3c00260,HAN2020147117}, however such synergistic effects are not explored herein. 

Activation with respect to HER is often attributed to improved values of the H-adsorption free energy ($\Delta G_\mathrm{H}$) on both the basal plane and sulfur-terminated edges. However, theoretical works have shown that the nature of hydrogen evolution via \ce{Mo}-sites and \ce{S}-sites differs, as in the latter case a large Heyrovsky activation energy must be overcome \cite{doi:10.1021/jacs.5b03329,doi:10.1021/acs.jpcc.1c10436,D3CP00516J,D3CP04198K}. This challenges the common descriptor-based view as metal-bound hydrogen contributes to evolution more readily despite the near-thermoneutral adsorption in both cases. Thus, improved HER via activation of sulfur-sites is an unlikely mechanism. A further implication is that calculations of activation energy are necessary to predict the HER performance at different dopant sites. 

Hereby, we investigate the consequences of transition/noble metal impurities (Co, Ni, Pt) on the HER kinetics through theoretical reaction modelling at the atomic scale. Our starting point is to consider the active sites of the undoped edge- and basal-oriented \ce{MoS2}, which are respectively the sulfur-depleted \ce{Mo}-edge (\ce{Mo}$_0$) and \ce{S}-vacancies, and their modification via doping. Overall, the results demonstrate how the presence of impurities can lead to reduction in the overpotential, and provide insight into the role of dihydrogen intermediates.


\section*{Methods}
\label{methods}
The electronic structure calculations were performed at the level of density functional theory (DFT) within the plane-wave formulation implemented in the Vienna ab initio simulation package (VASP) \cite{PhysRevB.49.14251,KRESSE199615, PhysRevB.54.11169}, with core electrons described by the projector-augmented wave approach \cite{PhysRevB.59.1758}. All calculations were spin-polarized, with plane wave basis sets up to $400$~eV and $3\times3\times1$ Monkhorst-Pack sampling of the Brillouin zone. Valence electrons for transition metals include the outer $s$- and $d$-electrons. The revised Perdew-Burke-Ernzerhof exchange-correlation functional \cite{PhysRevB.59.7413} was used, along with the D3 dispersion corrections \cite{doi:10.1063/1.3382344}. Transition states for elementary reaction steps were approximated by first-order saddle points on the potential energy surface, obtained by the climbing-image nudged elastic band method \cite{doi:10.1063/1.1329672}. Minima and saddle points were optimized with force convergence criteria of $0.02$~eV/\AA\ and $0.05$~eV/\AA, respectively. 

The electrolyte is represented by an explicit Eigen cation water cluster, as well as an implicit polarizable continuum model as implemented in VASPsol \cite{doi:10.1063/1.4865107,doi:10.1063/1.5132354}. For the final kinetic evaluations, the grand-canonical reaction and activation energy is evaluated so as to account for the constant electrode potential during the electrochemical reaction. This is done by allowing the number of electrons in the system to vary, implicitly tuning the electrode potential \cite{doi:10.1021/acs.jpcc.8b10046, D3CP00516J, D3CP04198K}. The grand-canonical energy is then defined as $\Omega = E_n + \delta n e \Phi$ where $E_n$ is the DFT energy with $n$ electrons, $\delta n = n-n_0$ is the difference in number of electrons from the neutral state, $e$ is the elementary charge and $\Phi$ is the electrode potential, given by the effective work function. In the following, we refer the electrode potential to the standard hydrogen electrode (SHE), defining $U=\Phi-\Phi_\mathrm{SHE}$ with $\Phi_\mathrm{SHE} = 4.43$~V. During the grand-canonical evaluation, the saddle point structure representing the transition state is kept fixed.

The basal plane is modelled as a single monolayer, represented by a periodically repeating $5\times5$ \ce{MoS2} unit cell (75 atoms excluding adsorbates and solvent species). The edge model consists of alternating layers terminated with $50$\,\% and $0$\,\% \ce{S}-coverage on the \ce{S}- and \ce{Mo}-edges, respectively, resembling a slab of vertically aligned 2H-\ce{MoS2} sheets. Each layer is $3\times4$ \ce{MoS2} unit cells for reaction calculations, and $5\times4$ for formation energy and adsorption calculations (64 and 112 atoms per simulation cell). Single transition metal atoms are introduced as \ce{Mo}-substitutional impurities to these model systems, yielding a basal plane doping concentration of $4$\,\% in terms of \ce{Mo}-substitution, and $25$\,\% or $100$\,\% substitution of the \ce{Mo} edge. The inset in Figure \ref{fig:F1} shows a top-view of the two simulation cells.

Formation energy calculations are performed via the cohesive energy
\begin{equation}
    E_c = E_{\mathrm{tot}} - \sum_i{E_i n_i},
\end{equation}
where the sum goes over unique atomic species $i$ with energy $E_i$ and of quantity $n_i$ in the simulation cell. 

The free energy of hydrogen adsorption in the gas phase is used for initial characterization of different adsorption sites, and is given by 
\begin{equation} 
    \Delta G = E_{n\mathrm{H}}-E_{(n-1)\mathrm{H}} - \frac{1}{2} E_{\mathrm{H}_2}+ \Delta E_\mathrm{zpe} - T\Delta S,
\end{equation}
where $E_{n\mathrm{H}}$ is the DFT energy of a state with $n$ adsorbed \ce{H}-atoms and $E_{\mathrm{H}_2}$ is the DFT energy of the \ce{H2} molecule. $\Delta E_\mathrm{zpe}$ denotes the difference in zero-point energy upon adsorption, and is determined by vibrational analysis. The entropic contribution $T\Delta S$ is estimated by the standard entropy of the \ce{H2} molecule, effectively neglecting the entropy of the adsorbed state. 

The concentration of sulfur vacancies and hydrogen adatoms directly depends on the electrochemical conditions, specifically the $\mathrm{pH}$, electrode potential $U$ as well as the presence of sulfur species in solution. We assume that desulfurization occurs via electrochemical \ce{H2S} production, so that the sulfur chemical potential is
\begin{equation}
    \mu_\mathrm{S} = \mu_{\mathrm{H}_2\mathrm{S}} - 2(\mu_{\mathrm{H}^+} + \mu_{\mathrm{e}^-}),
\end{equation}
where the chemical potential of the proton-electron pair in solution is defined via the computational hydrogen electrode approach \cite{doi:10.1021/jp047349j}:
\begin{equation}
    \mu_{\mathrm{H}^+} + \mu_{\mathrm{e}^-} = \frac{1}{2}\mu_{\mathrm{H}_2} - eU + k_\mathrm{B}T \ln a_{\mathrm{H}^+},
\end{equation}
where $\mathrm{pH}=-\log_{10} a_{\mathrm{H}^+}$. As for \ce{H2}, $\mu_{\mathrm{H}_2\mathrm{S}}$ is given by the DFT energy, zero-point energy and standard entropy. Acidic conditions ($\mathrm{pH}=0$) and a partial \ce{H2S} pressure of $p_{\mathrm{H}_2\mathrm{S}}/p^\circ_{\mathrm{H}_2\mathrm{S}} = 10^{-8}$ are assumed. These definitions of sulfur and proton-electron pair chemical potentials are used to evaluate relative energies of hydrogenated sulfur vacancies on the basal plane, as a function of electrode potential. 

A kinetic model is used to evaluate the relative reaction rates of the different configurations, and construct simulated polarization curves. We consider one site at a time, and separately evaluate the Volmer-Tafel and Volmer-Heyrovsky pathways. Assuming constant concentration of solvated species, the state space is given by the possible surface hydrogen configurations for the given site (0 or 1 \ce{H} for the Heyrovsky pathway, 0, 1 or 2 \ce{H} for the Volmer-Tafel pathway). Transitions between different microstates occur via the elementary Volmer, Heyrovsky and Tafel steps, and reverse reactions involving \ce{H2} are neglected, i.e. \ce{H2} is considered to be removed from the local system as soon as Tafel or Heyrovsky combination has occurred. A general transition rate from microstate $a$ to $b$ via path $s$ is given by $r^s_{ab} = \theta_a k^s_{ab}$, with $\theta_a$ being the occupation of state $a$. Within transition state theory, the associated rate constant takes the form
\begin{equation}
    k^s_{ab} = \frac{k_\mathrm{B}T}{h} e^{-\Omega^{\ddag(s)}_{ab} / k_\mathrm{B} T},
\end{equation}
where $k_\mathrm{B}$ is the Boltzmann constant, $h$ is the Planck constant, $T$ is temperature, and $\Omega^{\ddag(s)}_{ab}$ is the grand-canonical activation energy, i.e. the energy of the transition complex along $s$ connecting the $a$ and $b$ minima, referred to that of the state $a$. $T=300$~K is used in rate calculations. To account for rearrangement of the solvent beyond those included in the simulation, a minimum limit of $0.2$~eV is enforced for the Volmer and Heyrovsky forward/reverse barriers. The total transition rate from $a$ to $b$ is then a sum over the available pathways, $r_{ab} = \sum_s r^s_{ab}$.  Solving for the steady state condition, where the occupation of each state is constant, the steady state current density due to electron transfer in the forward ($+$) and reverse ($-$) Volmer ($\mathrm{v}$) and Heyrovsky ($\mathrm{h}$) steps is given by \begin{equation}
    j = -\frac{e}{A} \sum_{ab}\left( r_{ab}^\mathrm{v(+)} - r_{ab}^\mathrm{v(-)} + r_{ab}^\mathrm{h(+)}\right),
\end{equation}
with the elementary charge $e$ and effective area per site $A$. $j$ is voltage-dependent via the grand-canonical activation energies, and with this knowledge one can construct theoretical polarization curves. The area per site is here given by the number of sites (1 for basal plane, 1-4 for edges) and lateral dimensions of the chosen simulation cell. Our goal is not a direct comparison of the absolute magnitude of the current density with experiments, but nonetheless we should note that the edge content may vary considerably in experiments, depending on the sample morphology. The \ce{Mo}$_0$ edge content of this simulation cell is ca. $0.8$~nm/nm$^2$, which is of comparable magnitude to experimental values \cite{doi:10.1126/science.1141483}. As the current density is proportional to the site density, the exponential nature of the polarization curve somewhat mitigates the importance of the chosen area on determining the overpotential. When discussing the theoretical overpotential $\eta$ in the following, we refer to the magnitude of the negative electrode potential at which $j$ exceeds the arbitrary threshold of $10$~mA/cm$^2$. 


\section*{Results and Discussion}
\label{results_discussion}

\begin{figure}[!b]
    \centering
    \includegraphics[width=1\linewidth]{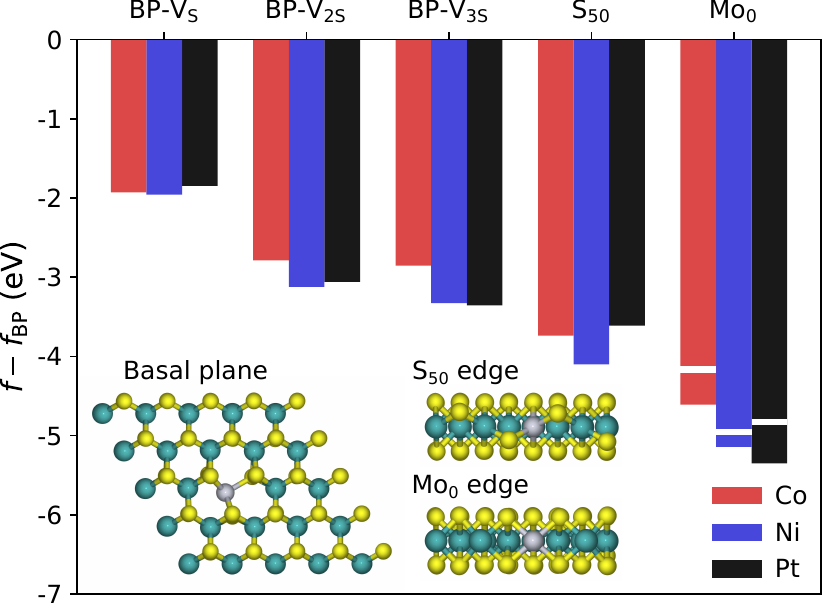}
    \caption{Relative formation energy of \ce{Mo}-substitutional doping in \ce{MoS2}. For all cases, sulfur-vacancies (V$_\mathrm{S}$, V$_{\mathrm{2S}}$, V$_{\mathrm{3S}}$) significantly stabilize the dopant atom relative to the pristine basal plane (BP, zero-level), but the lowest energy configuration is on the sulfur-depleted \ce{Mo}$_0$ edge. Inset shows examples of doping on the basal plane and edges. Horizontal lines for \ce{Mo}$_0$ indicate the formation energy per impurity atom in the case of complete edge substitution.}
    \label{fig:F1}
\end{figure}

Considering first the relative free energy of the impurity formation (Figure \ref{fig:F1}), we find that substitution on the edges is the most favored thermodynamically for all three metals, and especially for the \ce{Mo}$_0$ edge. Thus, it is more likely to occur on edge-locating \ce{Mo}. Further, interaction with sulfur-vacancies stabilizes the basal plane impurity. Defining chemical potentials from the respective bulk metal phases, the absolute basal plane substitution energies (zero-levels in Figure \ref{fig:F1}) are $3.8$ (Co), $4.3$ (Ni), and $4.5$~eV (Pt). However, conclusions regarding absolute stability warrant a more thorough thermodynamic study, considering other possible reference phases in appropriate conditions. Hence we consider the relative formation energy to evaluate the different sites for each impurity metal. In this regard, \ce{Co}, \ce{Ni}, and \ce{Pt} all show similar behavior. Based on these energies one might expect a significant edge-substitution at thermodynamic equilibrium, if no phase separation occurs. 
As indicated by horizontal lines in Figure \ref{fig:F1}, the formation energy per impurity atom remains low in the case of complete edge substitution, suggesting that a complete filling of the \ce{Mo}$_0$ edge is more favorable than occupying other sites. Moving forward, we assume from this that the doped \ce{Mo}$_0$ edge is present in the alternating vertical layer model. Note that the selection of the \ce{Mo}$_0$ and \ce{S}$_{50}$ edges qualitatively corresponds to sulfur-poor conditions, and that the selectivity of the substitution in general depends on the chemical potential of sulfur, as well as the impurity element \cite{SCHWEIGER200233,KIBSGAARD2010195,LAURITSEN2007220,Hakala2017}. 

The differences in substitution energy between basal plane and edges are in good agreement with those of Ref. \citenum{Hakala2017}, where it was also found that doping of the \ce{Mo}$_0$ edge was significantly more favorable than with $50$\,\% or $100$\,\% terminating \ce{S}-coverage. The presence of impurity atoms on the \ce{Mo} edge would then stabilize the sulfur-depleted configuration to some degree.

\subsection*{Doping of vacancies on basal plane}

\begin{figure}[h]
    \centering
    \includegraphics[width=1\linewidth]{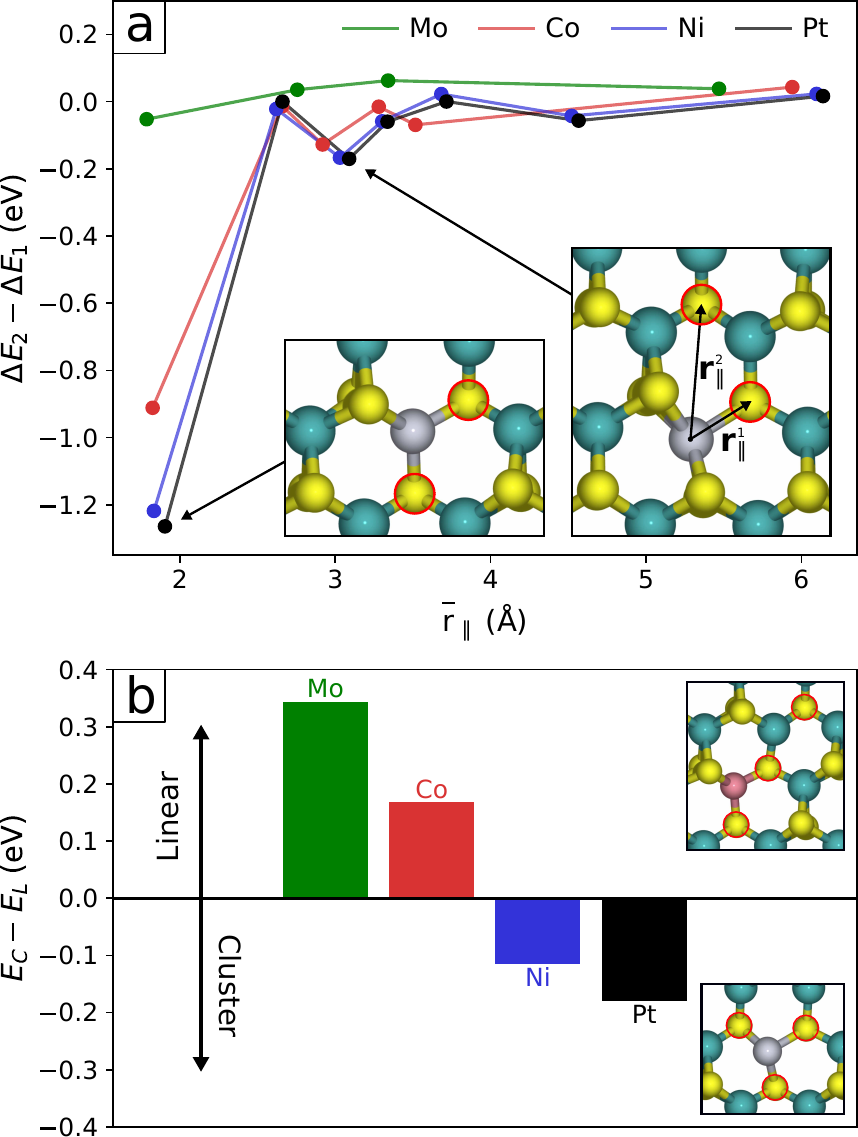}
    \caption{Doping effect on vacancy configurations. \textbf{a)} Second \ce{S}-vacancy formation energy, relative to the cost $\Delta E_1$ of the first vacancy formation on pristine \ce{MoS2}. The presence of transition metal dopants stabilizes neighboring vacancy configurations (smallest average lateral distance $\bar{r}_\parallel$). On undoped \ce{MoS2}, the neighboring configuration is only modestly favored, while the dopant-stabilizing effect is on the order of $1$~eV. \textbf{b)} Linear trimer vacancy distribution is favored in undoped \ce{MoS2} and with \ce{Co}-doping. \ce{Ni} and \ce{Pt} enable formation of clustered vacancy configurations, leaving the central (dopant) metal atom under-coordinated.}
    \label{fig:F2}
\end{figure}

Looking further into the interaction between \ce{S}-vacancies and the impurity metal, we note first that the single vacancy is stabilized by almost $2$~eV if situated next to an impurity atom in comparison with the undoped basal plane. This is in good agreement with an earlier theoretical work \cite{C9NR10702A}. In thermodynamic equilibrium the impurities will then be accompanied by \ce{S}-vacancies. Considering higher levels of S-deficiency and interaction between vacancies, neighboring vacancy pairs are weakly favored over dispersed configurations by ca. $0.1$~eV on the undoped \ce{MoS2} basal plane, see Figure \ref{fig:F2}a. 
In the presence of impurity atoms, however, the neighboring configuration is strongly favored by $\sim1$~eV for all dopants. Thus, the impurities enable neighboring vacancy configurations to a larger degree. Further, for the third vacancy a cluster-like configuration is favored for \ce{Ni}- and \ce{Pt}-doping, while \ce{Co}-doped and undoped \ce{MoS2} favor linear vacancy distributions, see Figure \ref{fig:F2}b. The cluster triple vacancy configuration fully exposes a metal atom and is likely of interest for HER. Overall, the vacancies and impurity atoms are co-confining. 

\begin{figure}[h]
    \centering
    \includegraphics[width=1\linewidth]{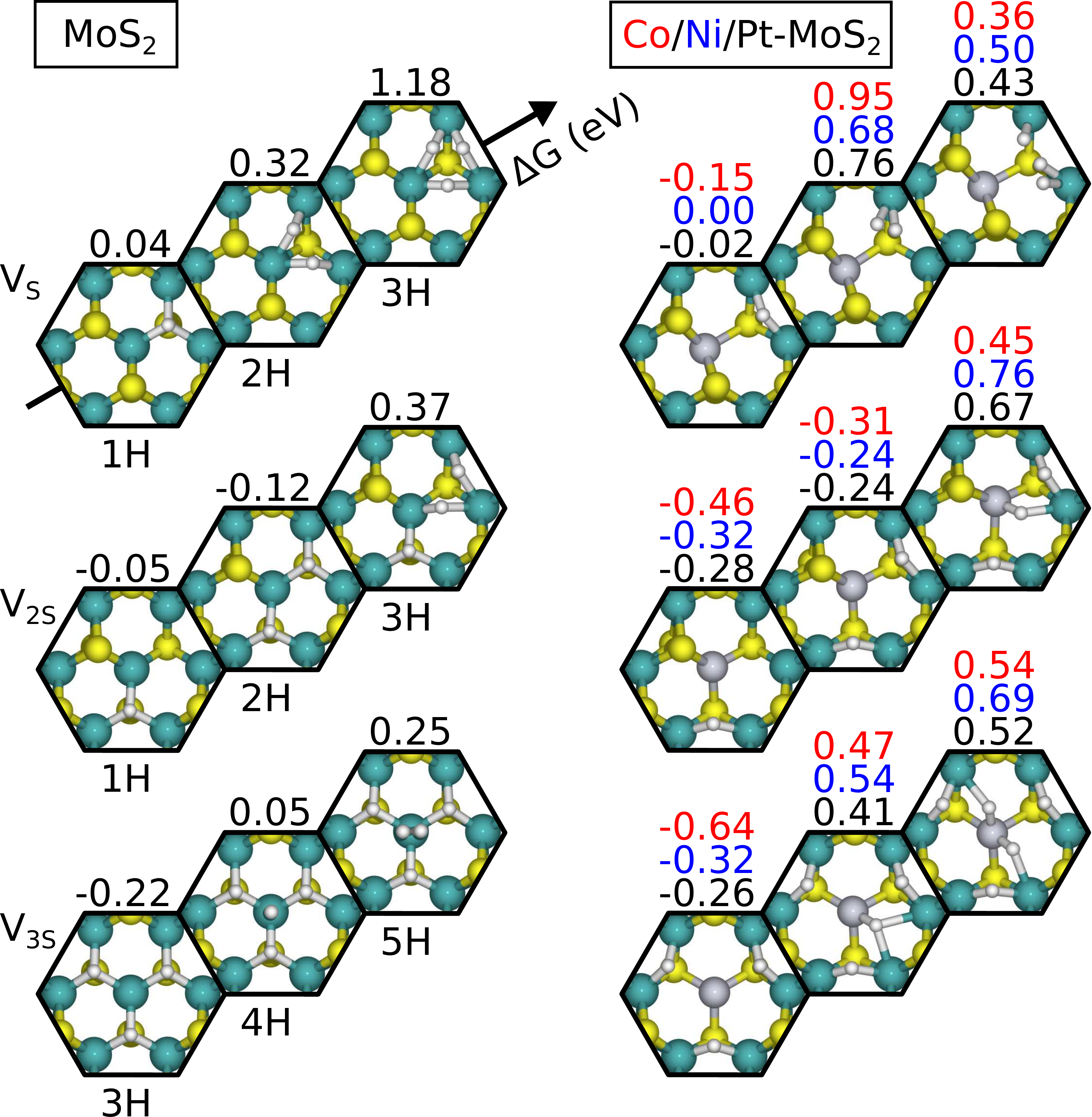}
    \caption{Adsorption configurations for the dopant-vacancy combinations. The differential free energy ($\Delta G$) with respect to the previous configuration is indicated. For the first image the reference is the preceding coverage (0\ce{H} for V$_\mathrm{S}$ and V$_\mathrm{2S}$, and 2\ce{H} for V$_\mathrm{3S}$). The local geometry is similar for the doped systems and is here represented by the \ce{Pt}-system. In most cases the doping leads to adsorption outside of the thermoneutral regime.}
    \label{fig:F3}
\end{figure}

On undoped \ce{MoS2}, hydrogen atoms can adsorb in the single vacancy site with a near-zero change in free energy. The multiple-vacancy configurations and impurity atoms modify the adsorption, as shown in Figure \ref{fig:F3}. Notably, \ce{Ni} and \ce{Pt} do not modify adsorption onto the single vacancy (V$_\mathrm{S}$) significantly, while \ce{Co} leads to a more favorable adsorption. In the doped vacancy, the second adsorption is significantly less favorable, but the adsorption configuration consists of \ce{H2}* adsorbed onto a \ce{Mo}-atom, which suggests a possible Volmer-Tafel pathway. The importance of dihydrogen intermediates has been established for single-atom catalysts \cite{doi:10.1021/jacs.1c10470}, and similar trends may be relevant on the local doping-induced structures considered in this work. In this specific case the intermediate seems too unstable to support an efficient pathway.
In the double-vacancy (V$_\mathrm{2S}$), adsorption becomes stronger by doping, and these configurations are likely not of interest for hydrogen evolution. In the clustered triple vacancy configuration (V$_\mathrm{3S}$), undoped \ce{MoS2} has a near-thermoneutral adsorption onto the top-site once the more favorable vacancy sites are filled. The presence of an impurity atom makes the top site less favorable. A bridge-configuration is preferred instead, but is still high in energy. Notably, a fifth hydrogen atom may adsorb (unfavorably) onto the undoped top site to form a dihydrogen complex. On the doped structures such a complex would dissociate, but even then the adsorption is too endothermic. Overall, the doping seems to bring the vacancy sites out of the thermoneutral regime for hydrogen adsorption. 

\begin{figure}[!h]
    \centering
    \includegraphics[width=1\linewidth]{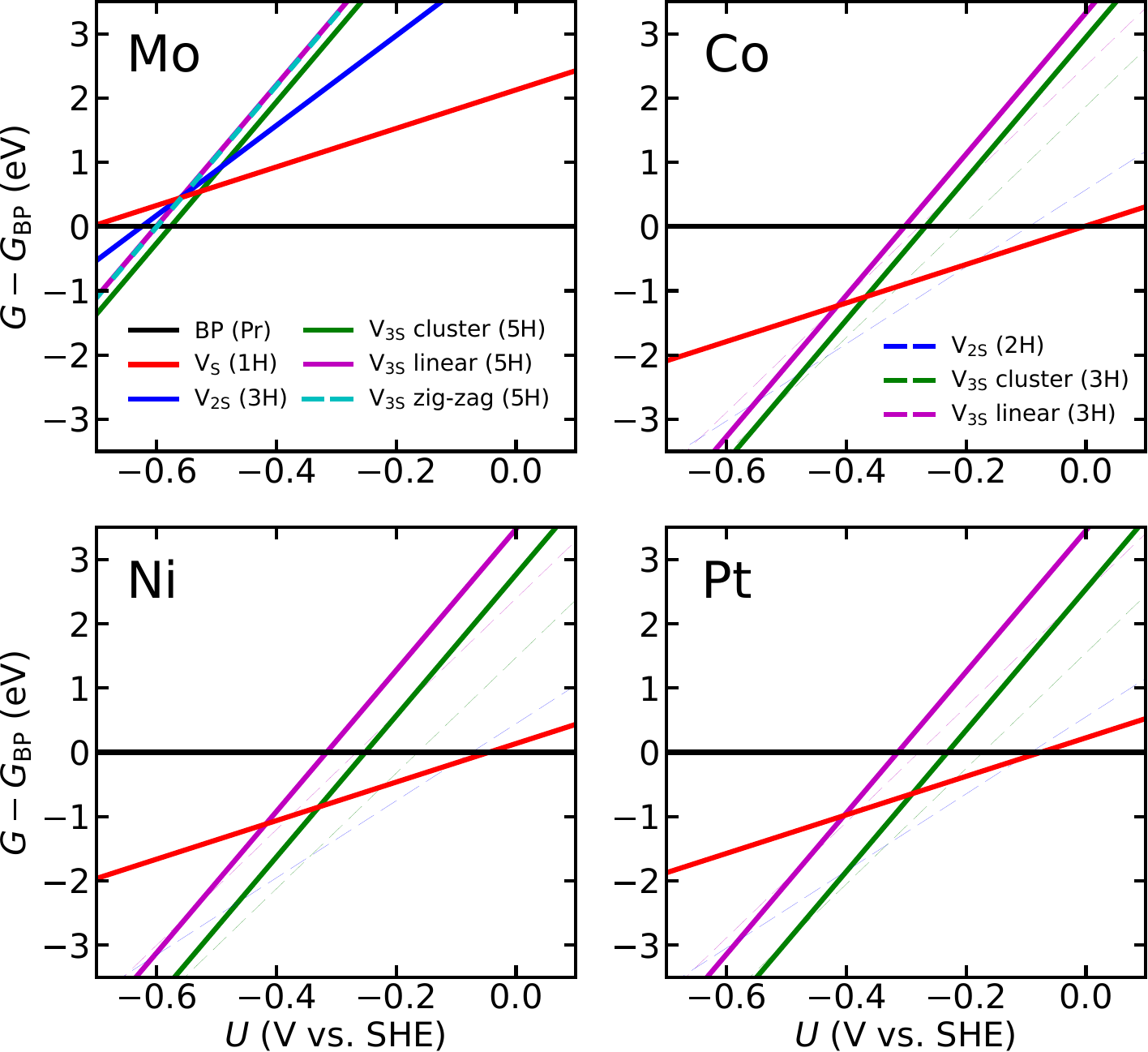}
    \caption{Free energy of hydrogenated vacancy configurations under electrochemical conditions at $\mathrm{pH} = 0$ and $p_{\mathrm{H}_2\mathrm{S}}/p^\circ_{\mathrm{H}_2\mathrm{S}} = 10^{-8}$. The reference (black line, zero) is the undoped or doped (but otherwise pristine) \ce{MoS2} basal plane. Under these conditions, the cluster vacancy (green line) is energetically preferred over the linear or zig-zag configuration in all systems, due to the greater \ce{H}-binding capacity. The cluster vacancy is the most stable configuration below ca. $-0.35$ to $-0.40$~V (doped) or $-0.55$~V (undoped).  Non-competing (higher energy in this region) hydrogen coverages for each vacancy configuration are not shown. }
    \label{fig:F4}
\end{figure}

Once the \ce{MoS2} system is subjected to electrochemical HER conditions, the potentially stabilizing effect of hydrogen adsorption is crucial in determining which vacancy configurations are formed. Under desulfurizing conditions ($\mathrm{pH}=0$ and $p_{\mathrm{H}_2\mathrm{S}}/p^\circ_{\mathrm{H}_2\mathrm{S}} = 10^{-8}$), Figure \ref{fig:F4} shows the free energy of hydrogenated vacancy configurations in doped and undoped \ce{MoS2}. Single and double vacancy formation is favored at small applied negative voltages for the doped systems, while the triple vacancy is preferred below ca. $U=-0.35$ to $-0.40$~V. In undoped \ce{MoS2}, the fivefold hydrogenated triple vacancy configuration surpasses the pristine basal plane around $-0.55$~V. At larger negative potentials, higher vacancy concentrations would be preferred, but importantly we note from these results that due to favorable hydrogen adsorptions, the clustered triple vacancy is favored over the linear or zig-zag variants in all systems. In the two latter cases, the adsorptions are higher in energy, more similar to those on the double vacancy. 
Given that the system can equilibrate under these conditions, we thus expect the clustered triple vacancy to be present (and favored) even in undoped \ce{MoS2}, and going forward we will consider the cluster vacancy for all systems. Tsai et al. \cite{Tsai2017} observed clustered vacancy configurations after electrochemical desulfurization of basal-oriented (undoped) \ce{MoS2}. Further, their observed onset in reduction of the \ce{S}:\ce{Mo} ratio between $U=-0.5$ and $U=-0.6$~V vs. RHE  coincides well with the stable region of V$_\mathrm{3S}$ for the undoped system in Figure \ref{fig:F4}.

\begin{figure}[t]
    \centering
    \includegraphics[width=1\linewidth]{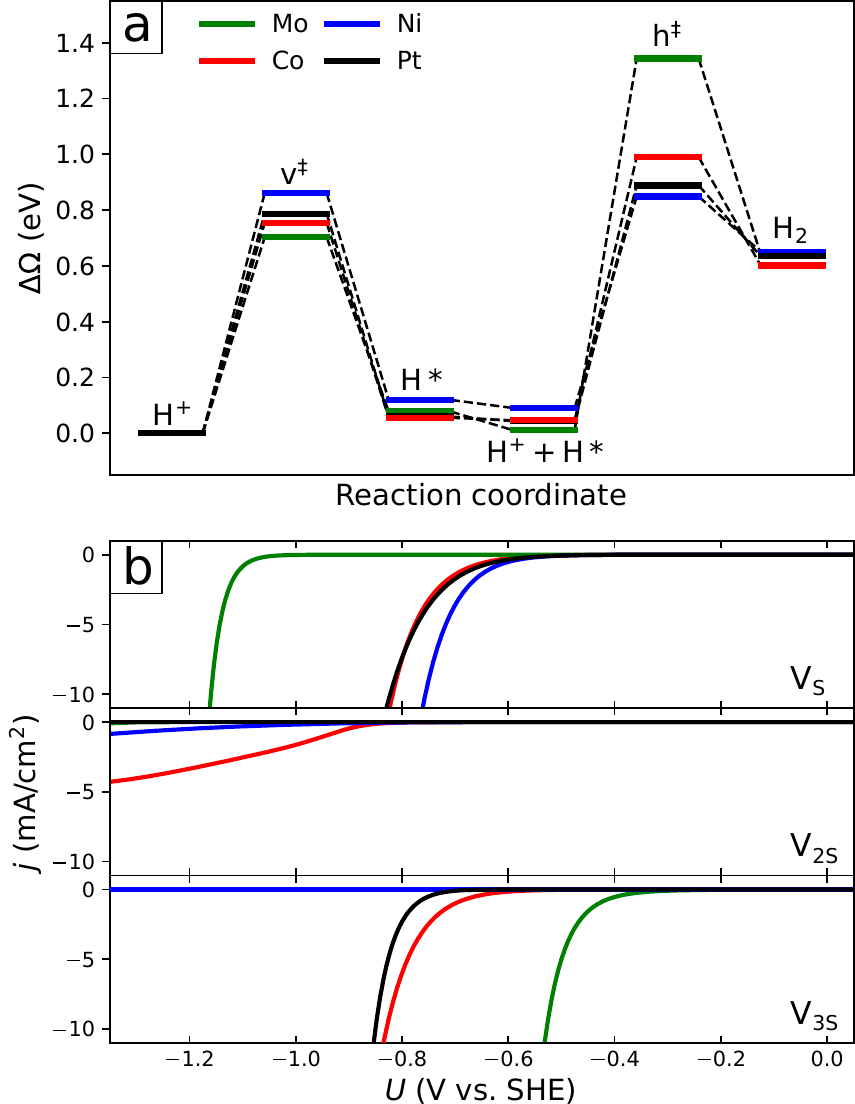}
    \caption{\textbf{a)} Grand canonical energy diagram for the Volmer-Heyrovsky process on the single vacancy configuration (V$_\mathrm{S}$) at $U=0$~V. Doping reduces the required Heyrovsky barrier, despite minimal changes in the \ce{H}-binding energy. \textbf{b)} Theoretical polarization curves for the basal plane vacancy configurations, with $4$\,\% Mo-substitution and $2-6$\,\% S-deficiency. All dopants activate the single vacancy site with respect to that in undoped \ce{MoS2}. \ce{Co} and \ce{Pt} display similar activity at both low (V$_\mathrm{S}$) and high (V$_\mathrm{3S}$) vacancy levels, while \ce{Ni} is deactivated in the latter case.}
    \label{fig:F5}
\end{figure}

Next we consider the single, double and triple (cluster) vacancies, and obtain the grand-canonical reaction- and activation energies via reaction modelling as outlined in the Methods section.  For the single-vacancy, a Volmer-Heyrovsky mechanism via the first hydrogen is preferred, despite the Tafel-relevant geometry upon double adsorption. The energy diagram for this mechanism (see Figure \ref{fig:F5}a), shows that the Heyrovsky barrier is significantly reduced by doping. The simulated polarization curves are presented in Figure \ref{fig:F5}b. As expected from the energy diagram, all dopants activate the single-vacancy with respect to the undoped case, reducing the required overpotential by $0.3-0.4$~V. 

On the double vacancy configuration, the Volmer-Heyrovsky pathway is preferred, but only \ce{Co} shows any significant activity. The Heyrovsky barriers are larger than in the single vacancy due to the more favorable adsorption, leading to the current onset at a more negative potential. Interestingly, the Heyrovsky barrier becomes smaller than the Volmer barrier at larger negative potentials, and continues to scale while the Volmer barrier stagnates. This leads to the flat shape of the \ce{Co}-\ce{MoS2} polarization curve, as the (more) rate-determining step is not being significantly reduced. For reference, calculations of the Tafel mechanism on these doped double-vacancy configurations showed that the barrier for Tafel combination is very large ($2$~eV upwards) due to repulsion from the impurity atom, rendering this mechanism irrelevant. The double vacancy thus strictly inhibits evolution in both doped and undoped cases.

\begin{figure}[t]
    \centering
    \includegraphics[width=1\linewidth]{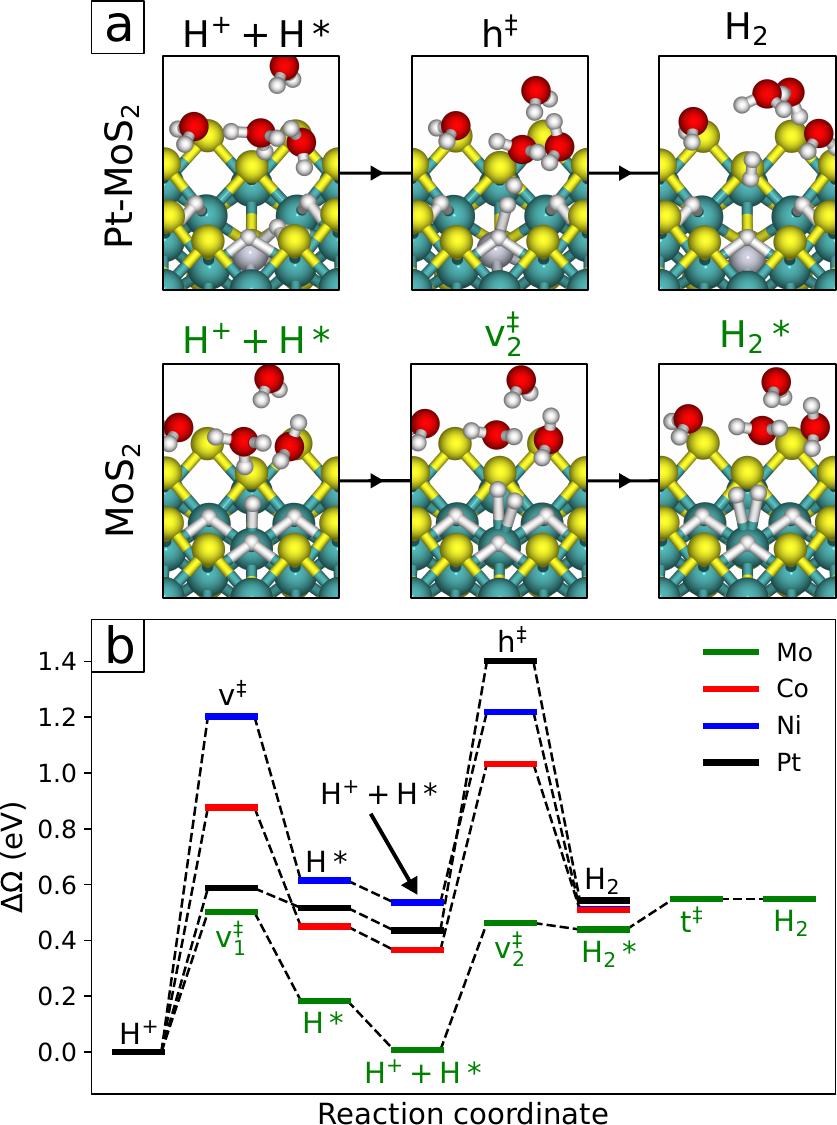}
    \caption{\textbf{a)} Reaction path on the doped and undoped cluster vacancy (V$_\mathrm{3S}$) after the first Volmer step. A second Volmer step is possible in the undoped case. \textbf{b)} Grand canonical energy diagram at $U=0$~V for the cluster vacancy configuration. Doping disables the path that proceeds via an \ce{H2}* intermediate, and favors instead direct Heyrovsky combination with a larger barrier. }
    \label{fig:F6}
\end{figure}

For the triple vacancy, different mechanisms are supported in doped and undoped systems. A Volmer-Volmer-Tafel mechanism via a dihydrogen intermediate is possible in the undoped case, while the endothermic \ce{H}-adsorption of the doped systems favors a direct Volmer-Heyrovsky mechanism (see Figure \ref{fig:F6}a). The corresponding energy diagrams are shown in Figure \ref{fig:F6}b, where the required activation energy is significantly smaller in the undoped case. This leads to the low overpotential in Figure \ref{fig:F5}b. The \ce{Co}- and \ce{Pt}-systems display very similar overpotentials compared to the single vacancy case, while the \ce{Ni}-doped system is limited by a large Volmer-barrier and is deactivated in this configuration.
Importantly, formation of a dihydrogen complex enabled a more efficient reaction path, which is further explored in the following section on edges. 

The improved kinetics of the clustered triple vacancy on undoped \ce{MoS2} has some important implications. Due to the large overpotential required on the undoped single vacancy, triple vacancies would form before this point according to the analysis in Figure \ref{fig:F4}, indicating that the origin of basal plane activity is not only any \ce{S}-vacancy, but specifically the undercoordinated \ce{Mo}-atoms in threefold vacancy configurations. This is in contrast with the typical understanding that the single vacancy itself enables evolution due to its near-thermoneutral hydrogen binding. Comparing again with work by Tsai et al. \cite{Tsai2017}, the onset of reduced \ce{S}:\ce{Mo} atomic ratio around $U=-0.6$~V vs. RHE was accompanied by significantly increased HER activity. This aligns well with the understanding provided by our results, namely that the fully exposed \ce{Mo} atom is the active site on the basal plane. This configuration is generated by the same conditions that drive the hydrogen evolution, and connection with the initial vacancy concentration and distribution in the sample may be elusive. 

Doping deactivates the triple vacancy (relatively for \ce{Co} and \ce{Pt}, and completely for \ce{Ni}), which suggests that the basal plane is deactivated by doping overall. Doped single-vacancies are activated by a reduction of $0.3-0.4$~V in the overpotential, and are present at lower voltages than in the undoped case. However, they are only stable at very small voltages, before they are replaced by doped double- and triple vacancies, both of which inhibit hydrogen evolution compared to the undoped triple vacancy. Experimental work by Humphrey et al. \cite{C9NR10702A} found reduction in HER activity upon doping basal-oriented \ce{MoS2} with \ce{Co}, in agreement with these conclusions. Rather than deactivating the single vacancy, our results suggest that this is due to deactivation of the triple vacancy.

\subsection*{Doping of edges}

\begin{figure}[!h]
    \centering
    \includegraphics[width=1\linewidth]{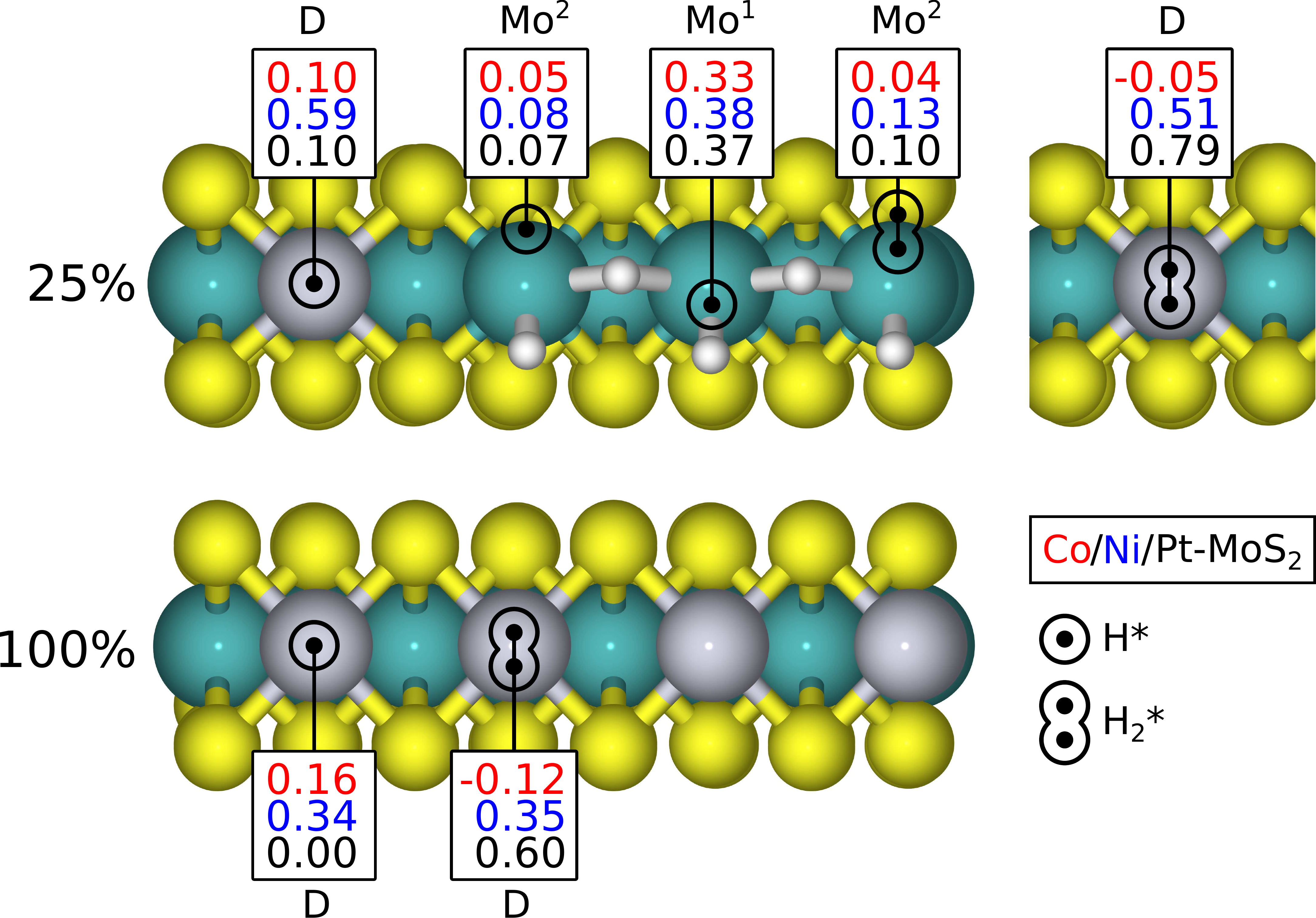}
    \caption{Differential free energy ($\Delta G$ in eV) of adsorption onto the H-saturated \ce{Mo}$_0$ edge where $25$\,\% and $100$\,\% of terminating \ce{Mo} is substituted by impurity atoms. For \ce{H2}* complex, the free energy is given for the entire adsorption complex. Note that $\Delta G > 0$ by default for the single H atom adsorption (1.25 monolayer saturated coverage).}
    \label{fig:F7}
\end{figure}

We have chosen to focus on the \ce{Mo}$_0$ edge, as this is both the most kinetically active edge \cite{D3CP04198K} as well as the most thermodynamically stable doping site. Two cases of edge doping levels are considered, $25$\,\% and $100$\,\%. At $25$\,\% edge doping, the equilibrium \ce{H}-coverage is reached at $1.25$ monolayers where \ce{H} is bound to \ce{Mo}-atoms, avoiding the slightly higher energy impurity site. From there on, the next adsorption can occur onto the impurity atom or onto one of the \ce{Mo} atoms, in most cases with a weakly positive free energy cost. The exception is adsorption onto the \ce{Ni}-atom, which is unfavorable by ca. $0.5$~eV, see Figure \ref{fig:F7}. 

At $100$\,\% edge doping, the first adsorption is (by default) endothermic for \ce{Co} and \ce{Ni}, and $0$~eV for \ce{Pt}. However, interestingly, another hydrogen atom can be adsorbed onto the same dopant site, forming an adsorbed \ce{H2}* complex. The adsorbed complex is stable for \ce{Co}, and for \ce{Ni} the second adsorption is nearly thermoneutral with respect to the first. On the contrary, the second adsorption is strongly unfavorable for \ce{Pt}. The same trend can be seen on the impurity atom at the $25$\,\% doping level. The affinity towards \ce{H}* and \ce{H2}* are thus qualitatively different for these metal impurities. We note that the trend of exothermic adsorption to form stable \ce{H2}* following an initial endothermic \ce{H}-adsorption is similar to that observed in single transition metal atoms supported on the \ce{MoS2} basal plane \cite{doi:10.1021/jacs.1c10470}. The \ce{H2}* complex can also form on \ce{Mo}-atoms, and in this case the adsorption is closer to thermoneutral also for the \ce{Ni}- and \ce{Pt}-systems. The formation of such \ce{H2}* species is relevant when considering the possible reaction pathways for HER. Such species with near-thermoneutral binding are prime candidates for efficient intermediates. This also illustrates the importance of considering coverage effects and exploring the adsorption configuration space prior to modelling HER itself. 

\begin{figure}[!h]
    \centering
    \includegraphics[width=1\linewidth]{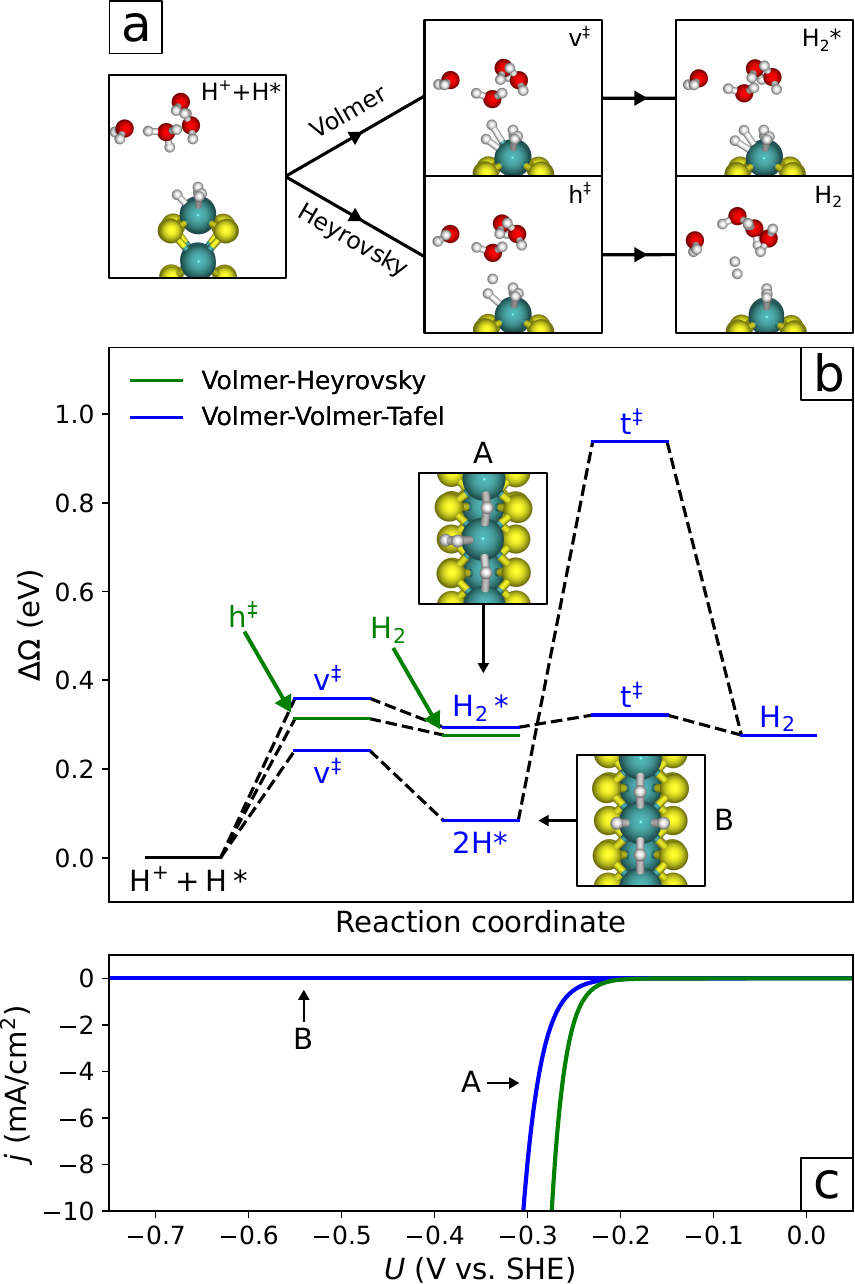}
    \caption{\textbf{a)} Attractive interaction between the solvated proton and the surface allows the Volmer and Heyrovsky reactions to proceed through nearly-similar TS configurations. \textbf{b)} Energy diagram at $U=-0.2$~V vs. SHE for the undoped \ce{Mo}$_0$ edge comparing the Volmer-Heyrovsky and Volmer-Volmer-Tafel pathways starting after the first Volmer step at $\theta_\mathrm{H} = 1.25$, which is the same for all cases. \textbf{c)} Resulting theoretical polarization curve from the above kinetics. In this case, the Volmer-Heyrovsky pathway proceeds at a lower overpotential.}
    \label{fig:F8}
\end{figure}

The possibility of further \ce{H}-adsorption leading to \ce{H2}* complexes warrants revisiting the undoped \ce{Mo}$_0$ edge, where the evolution was found to proceed through a Volmer-Heyrovsky pathway \cite{D3CP04198K}. Interestingly, the incoming proton interacts favorably with the surface in the corresponding Heyrovsky transition state (TS). This attractive interaction stabilizes the transition complex, and leads to geometrically similar TS for the Volmer and Heyrovsky steps which are also close on the potential energy surface (Figure \ref{fig:F8}a). From the Volmer perspective, TS is stabilized by favorable \ce{H-H} interaction in addition to the surface attraction. This also means that something resembling a \ce{H2}* complex near the surface is part of the reaction for both the Heyrovsky and Volmer processes. Depending on whether the energetics favor adsorption of the \ce{H2}* complex, \ce{H2} may release directly into solution or stay on the surface, (nominally a Heyrovsky or Volmer step). In the latter case, \ce{H2}* may then desorb in a Tafel-like process at a later point, completing HER. In the following we refer to these mechanisms as Volmer-Heyrovsky and Volmer-Volmer-Tafel, although the Tafel process only involves desorption, not \ce{H-H} combination. 

Figure \ref{fig:F8}b displays the grand canonical energy diagrams at $U=-0.2$~V for the undoped \ce{Mo}$_0$ edge at a hydrogen coverage $\theta_\mathrm{H} = 1.25$,  starting after the first Volmer step (identical for all cases). The \ce{H2}* intermediate is seen to proceed with a very small Tafel barrier after adsorption. However, the direct \ce{H2}-release via a Heyrovsky process is slightly preferred due to the smaller barrier, and proceeds at a lower overpotential, as shown in Figure \ref{fig:F8}c. Note that this distinction may be sensitive to computational specifics, e.g. choice of exchange-correlation functional and dispersion corrections. The Volmer-Tafel process through the \ce{H2}* intermediate (A) is also compared to that of the alternative \ce{2H}* intermediate (B). The \ce{2H}* configuration is more stable, but also requires a large Tafel barrier to combine, leading to detrimentally worse overall kinetics. The difference between these two Tafel paths illustrates that a more thermoneutral intermediate is not always associated with better kinetics, as possible pathways and barriers must be considered. The presence of these similar transition- and intermediate complexes suggests that the potential energy surface is quite flat in this immediate neighborhood of the configuration space, encompassing configurations of \ce{H2}* to \ce{H2} via water-surface complexation with comparable energies.

\begin{figure}[!t]
    \centering
    \includegraphics[width=1\linewidth]{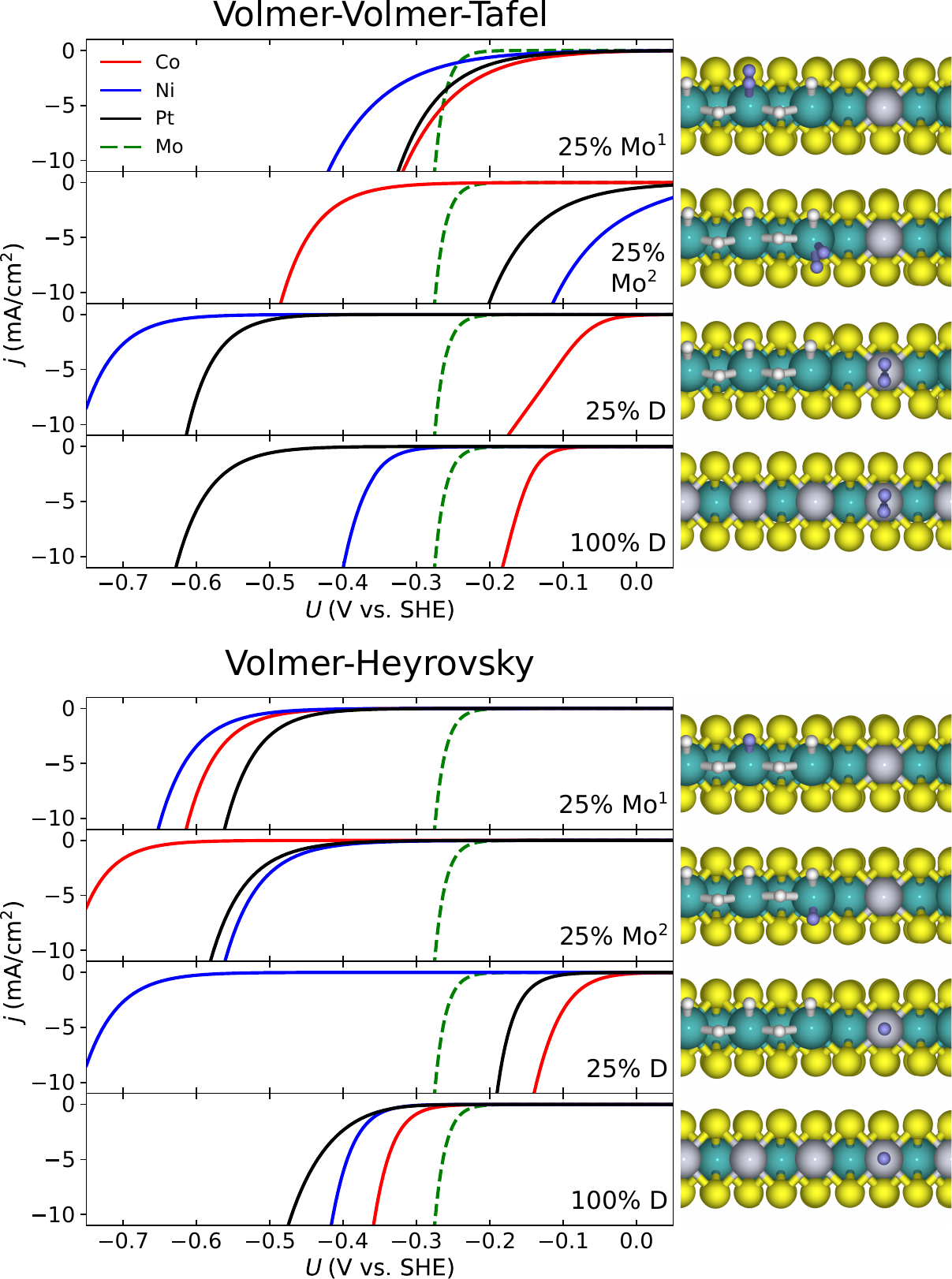}
    \caption{Theoretical polarization curves for the Volmer-Volmer-Tafel and Volmer-Heyrovsky pathways on edges with $25$\,\% and $100$\,\% impurity substitution. Green dashed line shows the undoped \ce{Mo}$_0$ reference. Panels show the structures for the relevant \ce{H}* or \ce{H2}* intermediates with hydrogen atoms partaking in the evolution colored in blue.}
    \label{fig:F9}
\end{figure}

Based on the findings above and considering the adsorption energetics, we map out the Volmer-Heyrovsky and Volmer-Volmer-Tafel pathways through the \ce{Mo}- and dopant-sites. The polarization curves are presented in Figure \ref{fig:F9}, where we consider hydrogen coverages immediately below and above the equilibrium coverage. The H atoms partaking in evolution are colored blue in the right panels. For both mechanisms, the kinetics on the \ce{Mo}-site farthest from the impurity atom (\ce{Mo}$^1$) are rather similar, which is to be expected. Closer to the dopant atom the behavior is more different. The Volmer-Heyrovsky mechanism is deactivated on both \ce{Mo}-sites, but activated on the dopant site for $25$\,\% of \ce{Co} and \ce{Pt} impurities. The \ce{Ni}-site is severely Volmer-limited due to the large adsorption energy. At $100$\,\%, all systems are moderately deactivated for Volmer-Heyrovsky, compared to the pristine Mo edge. The kinetics at $25$\,\% match the expectations set by adsorption energies, but this is not the case at $100$\,\%, where \ce{Pt} performs worse despite $\Delta G = 0$~eV. The \ce{Mo}-sites also perform worse despite near-zero adsorption energies, demonstrating that the interaction between the protonated water cluster and electronic states of the hydrogenated surface at the transition state can not be inferred from the hydrogen binding energy alone. The Volmer-Volmer-Tafel process is seen to proceed at a reduced overpotential via the dopant site for \ce{Co} ($25$\,\% and $100$\,\%), and via the neighboring \ce{Mo}-site for \ce{Ni} and \ce{Pt} ($25$\,\%), where the \ce{H2}* complex is not favored on the impurity site. 

While considering these different possibilities in the context of experiment, the first remark is that the \ce{Mo}$_0$ edge is significantly more active than the basal plane, with ca. $0.25$~V lower overpotential in the undoped cases. Therefore, if the edge-content in experiment is not negligible (even small amounts of edge-sites will affect the effective overpotential significantly), e.g. in polycrystalline samples, it is reasonable to assume that the observed doping effect is due to modification of the edge sites. Within this regime, moderate doping levels of \ce{Co}, \ce{Ni} and \ce{Pt} will enhance the edge activity. At high doping levels, \ce{Ni}- and \ce{Pt}-doping leads to deactivation while \ce{Co}-doping still turns out beneficial.

The activation due to \ce{Pt} by Deng et al. \cite{C5EE00751H} and \ce{Co} by Lau et al. \cite{C8SC01114A} leads to shifts in the overpotential of similar magnitude as those found here for the $25$\,\% \ce{Mo}$_0$ edge. 
The deactivation due to \ce{Ni} by Lau et al. is also similar to what we find for the $100$\,\% edge substitution. Deng et al. find also weak activation by \ce{Co}, yielding a slightly larger overpotential than for \ce{Pt}-\ce{MoS2}, while we find the opposite trend in the theoretical activation. The reason for this is not clear, but we should note that fixed wt\,\%-doping constitutes different degrees of atomic substitution for \ce{Co}/\ce{Ni} and \ce{Pt}. Due to the difference in atomic mass, the distribution of \ce{Pt}-impurities will be more than three times as dilute. Deng et al. use a lower doping level ($1.7$~wt\,\%) than Lau et al. ($3.0$~wt\,\%) and see a less pronounced \ce{Ni}-deactivation. We speculate that this difference is related to the degree of edge-substitution, as it is consistent with the theoretical understanding developed in this work (activation for partial edge substitution, deactivation upon complete edge substitution). The experimental work that most resembles the edge model used here is that of Wang et al. \cite{Wang2015} where vertically aligned \ce{MoS2} layers were doped with 3d metals \ce{Fe} through \ce{Cu}. The \ce{Co}-doped sample contained ca. $22$ at\,\% \ce{Co} on the edge, decaying with depth, and all dopants resulted in overpotential reduction on the order of $0.1$~V, in excellent agreement with our findings for the $25$\,\% edge substitution. Rather than activation of the \ce{S}-edge, our results indicate that this activation can be explained by improved kinetics on the \ce{Mo}$_0$-edge. 

It should be noted that any systematic errors due to choices in the theoretical model, e.g. the implicit solvation scheme, explicit interface description, exchange-correlation functional or other computational parameters, may manifest as shifts in predicted barriers and reaction rates. We assume that these errors are similar across the studied systems, so that relative comparison is valid. Comparison with experiment suggests that the absolute rates are also of reasonable magnitude, but we must acknowledge the possibility of errors due to neglected effects, as well as partial cancellation of these.

Unlike the edges, our results show that the 2H-\ce{MoS2} basal plane is deactivated by \ce{Mo}-substitutional doping. This doping mechanism cannot therefore account for the very low overpotentials observed in experiments. However, we can not disregard the possibility that the general basal plane may be activated by means of other mechanisms, notable examples being transition to the 1T phase or anchored single-atom impurities. 

Both the basal plane \ce{S}-vacancies and the \ce{Mo}$_0$ edge involve \ce{Mo}-bound hydrogen with $\Delta G_\mathrm{H}\approx 0$, but the much faster kinetics on the edge (and the significant improvement from single to triple vacancy) emphasizes the significance of exposing undercoordinated metal atoms, and suggests that the resulting hydrogen binding is of different nature on the respective sites. The low coordination of the edge enables formation of the metal-bound dihydrogen complex. Such affinity towards dihydrogen chemisorption is typically associated with transition metal complexes \cite{doi:10.1021/ja00314a049,doi:10.1021/ar00147a005,doi:10.1021/ar00172a001}, as \ce{H2} on metal surfaces tends to either dissociate or physisorb \cite{CHRISTMANN19881} (dihydrogen may however bind to defects such as ridges \cite{doi:10.1021/acs.jpcc.8b10046,Gudmundsdottir12,Gudmundsdottir13} or form as transient states during Tafel combination \cite{doi:10.1021/acscatal.1c00538}). This indicates that the \ce{Mo}$_0$ edge represents a middle ground between undercoordination, enabling chemistry resembling a coordination compound, and maintaining the good electron transport and stability of a metallic surface, as can be a problem with e.g. supported single atom catalysts. The hapticity of \ce{H2} ligands on transition metal complexes is largely governed by $\pi$-backdonation from metal $d$-orbitals into the antibonding $\sigma^*_\mathrm{H-H}$ orbital, and equivalent mechanisms likely determine the \ce{H2}* binding energy on the doped edges.  From the few dopants studied herein one can at least note that dihaptic binding on the edge is less favored for dopants with a higher number of valence $d$-electrons (\ce{Co}, \ce{Ni} and \ce{Pt} being respectively $3d^7$, $3d^8$ and $5d^9$), though more in-depth analysis would be insightful in this regard. As the Kubas interaction fundamentally defines the energy landscape of the \ce{H2}*$\leftrightarrow$\ce{2H}* transition, tuning this interaction may be an important tool also in optimizing electrocatalyst performance.


\section*{Conclusions}
\label{conclusion}
	
The effect of transition/noble metal doping (\ce{Co}, \ce{Ni} and \ce{Pt}) on the activity of \ce{MoS2} towards HER was studied by performing grand-canonical DFT simulations (incl. solvent description) and theoretical reaction modelling which enabled construction of theoretical polarization curves. For the undoped basal plane, the active site responsible for intrinsic activity was found to be the central \ce{Mo} atom in clustered threefold \ce{S}-vacancies. Despite the single- and double-vacancies allowing near-thermoneutral \ce{H}-adsorption, hydrogen evolution via a \ce{H2}* complex occurs through a Volmer-Volmer-Tafel pathway on the triple vacancy with a much lower overpotential ($\eta>1.1$~V vs. $\eta=0.52$~V). The clustered triple vacancy is also energetically favored by HER conditions. Impurity atoms interact favorably with \ce{S}-vacancies in mutually stabilizing configurations, enabling increased vacancy generation. However, \ce{H}-intermediates on the impurity atom are high in energy, and the resulting Volmer-Heyrovsky pathway requires a larger activation energy. Therefore, all dopants seemingly deactivate the basal plane. 

Like the undoped triple vacancy, the edges display moderate affinity towards surface-bound \ce{H2}* complexes in both doped and undoped cases. This enables an efficient Volmer-Volmer-Tafel process of evolution in which the \ce{H}* combination occurs directly in the second Volmer step, and also influences the Volmer-Heyrovsky process by stabilizing the transition complex via attractive surface interaction. At moderate doping levels ($25$\,\% edge substitution), evolution proceeds with a reduced overpotential ($0.1-0.2$~V) for all dopants. At full edge substitution, \ce{Ni} and \ce{Pt} are deactivated relative to the undoped \ce{Mo}$_0$ reference ($\eta = 0.27$~V), and only the Volmer-Volmer-Tafel process on the \ce{Co} site remains more active, seemingly due to the more stable dihydrogen intermediate, which may further be related to the valence $d$-electrons via the Kubas interaction. 

The large discrepancy in performance between sites on the basal plane and edges despite similar thermodynamics of the \ce{H}* intermediate illustrates the importance of explicitly including activation energy in kinetic estimations, as the free energy descriptor does not account for this difference. In experimental context, it suggests that observations of low overpotential in doped 2H-\ce{MoS2} are unlikely to be due to \ce{Mo}-substitutional doping of the basal plane, and rather due to modification of the \ce{Mo} edge. 

In summary, edge doping can greatly increase the HER activity of \ce{MoS2}, and the best results are achieved by partial (\ce{Co}, \ce{Ni}, \ce{Pt}) or full substitution (\ce{Co}) of the \ce{Mo}$_0$ edge. The different trends on \ce{MoS2} basal planes, low-, and high-level doped edges may help explain experimental observations of differing effect of transition and noble metal impurities.
Further, the role of dihydrogen intermediates was identified, contributing towards understanding the chemical picture of hydrogen evolution in these systems. The behavior of transition metal dichalcogenide edges presents an intersection between surface- and coordination chemistry which seems essential for their role as efficient catalysts. Further understanding of this regime may guide the design of active yet stable catalysts.


\clearpage
\section*{Acknowledgements}

The calculations were performed on resources provided by Sigma2 - the National Infrastructure for High Performance Computing and Data Storage in Norway, project No. NN9497K. JA acknowledges financial support from the Academy of Finland, project No. 322832 ‘‘NANOIONICS’’. 
HJ thanks Prof. Jens N{\o}rskov for numerous inspiring discussions and collaborations over the past 30 years, and acknowledges financial support from the Icelandic Research Fund, project No. 207283-053.


\section*{Conflict of Interest}

There are no conflicts to declare.

\begin{shaded}
\noindent\textsf{\textbf{Keywords:} \keywords} 
\end{shaded}


\setlength{\bibsep}{0.0cm}
\bibliographystyle{Wiley-chemistry}
\bibliography{bi}


\end{document}